\documentclass[aps,prl,preprint,groupedaddress,showpacs]{revtex4}

\begin{document}


\title{Paramagnetic and diamagnetic states in two-dimensional Josephson-junction 
arrays}


\author{Cinzia De Leo and Giacomo Rotoli\ddag}
\affiliation{Dipartimento di Energetica and \ddag Unit\'{a} di Ricerca INFM,\\ 
Universit\'{a} di L'Aquila, ITALY}


\date{\today}

\begin{abstract}
Many experiments on high-temperature superconductors have shown paramagnetic 
behavior when the sample is field cooled. The paramagnetism was first attributed to a 
$d$-wave order parameter creating $\pi$-junctions in the samples. However, the same effect 
was later discovered in traditional low-temperature superconductors and conventional 
Josephson-junction arrays 
which are $s$-wave. By simulating {\it both} conventional and mixed $\pi$/conventional 
Josephson-junction arrays we determine that differences exist which may be sufficient to 
clearly 
identify the presence (or absence) of $\pi$-junctions. In particular the $\pi$-junctions
cause a symmetry breaking providing a measurable signature of their presence in sample. 
\end{abstract}

\pacs{74.50+r, 74.72-h, 75.20 -g}

\maketitle


In recent years a significant discussion in the superconductivity field was devoted to the 
paramagnetic response in field-cooled samples, the so-called paramagnetic 
Meissner effect (PME),  \cite{Braunisch}. PME was first observed in high-$T_c$ ceramic 
materials \cite{Braunisch,Kirtley,AltriHTC} and then also in low $T_c$ samples \cite{Nbdisks,AltriLTC}. A mechanism to explain the effect was introduced by Sigrist and Rice 
\cite{SigrRice1} where PME was due to the presence of $\pi$-junctions in the multiple-connected network of junctions formed by the superconducting grains in the high-$T_c$ ceramic. $\pi$-junctions are Josephson junctions formed between superconductors with unconventional pairing which cause a $\pi$ shift in the phase-current relation \cite{VanH,Ryaza}. Unconventional pairing, such as a $d$-wave order parameter, was introduced to explain the properties of high-T$_c$ ceramic materials \cite{Annett}.
PME was taken as strong evidence for a $d$-wave order parameter \cite{SigrRice1,SigrRice2}.
Multiple-connectiveness and $\pi$-junctions were soon shown to give a
paramagnetic response in simulated Josephson junction arrays by Dominguez, Jagla and Balseiro \cite{Dominguez}.

The discovery of PME in conventional low-$T_c$ (LTC) samples shows that PME cannot always be attributed to $\pi$-junction because they are present only in $d$-wave superconductors (or in  
unconventional superconductor-ferromagnet systems \cite{Ryaza}).
New explanations such as a giant flux state \cite{GFS}, flux compression \cite{bla1} and 
surface states \cite{bla2} have been introduced. The common feature was that PME in LTC samples is described as non-equilibrium phenomenon. 

Very recently a new experiment was devised to test the connection between multi-connectiveness and PME in conventional systems. A square array of LTC junctions was field cooled and shown to be predominantly paramagnetic \cite{Nielsen}; this paper also proposed
a qualitative explanation for the effect. A subsequent analysis
performed by numerical simulations of the arrays confirmed the
presence of PME in LTC arrays \cite{Deleo}. On this basis the essential
ingredients of PME appears to be multiple-connectiveness rather than
the presence of $\pi$-junction or non-equilibrium effects. 

It is clear that simulated Josephson-junction arrays both with\cite{Dominguez} 
and without \cite{Deleo} $\pi$-junctions show PME. Here
we investigate by means of numerical simulations
the differences between the two cases and the true role of the
$\pi$-junctions. We also suggest an 
experimental signature of the presence of $\pi$-junctions in field-cooled array samples. 

The array is described by means of a full mutual inductance model 
similar to that used in ref. \cite{Deleo}. The main difference 
is the presence of some $\pi$-junction, i.e., Josephson junctions 
where supercurrent is proportioinal to $\sin(\varphi+\pi)$ instead of
$\sin\varphi$ (cf. Fig. 1). 
The dynamics of an NxN array are described by the following equations\cite{Deleo,Studies,ISEC2001}:

\begin{equation}
\frac{\beta _{L}}{2\pi }\;(1+\vec{\delta})\;\sin (\vec{\varphi}+\vec{\psi})
+\sqrt{\frac{\beta _{L}}{\beta _{C}%
}}\overrightarrow{\dot{\varphi}}+\overrightarrow{\ddot{\varphi}}=\hat{K}%
\hat{L}^{-1}\vec{m} .  \label{n-arr-eq}
\end{equation}

Here time is normalized to a cell frequency 
($\omega^{-2} ={L^{\prime} C}$), with $L^{\prime}$ the self inductance of the elementary loops 
formed of the array $C$ the junction capacitance. 
$\beta_C$ is the Stewart-McCumber parameter, $\beta_C = 2 \pi I_0 C / \Phi_0 G^2$ where $I_0$ 
is the (mean) Josephson critical current for the junctions in the array, $\Phi_0$ is the flux quantum and $G$ is the junction conductance. $\beta_L$ is equal to  $2 \pi L' I_0/ \Phi_0$. 
$\vec{m}$ represents the normalized loop
magnetization (in unit of $\Phi_0$) which depends on frustation $f$, phase vector $\vec{\varphi}$ and the initial distribution of flux quanta in the array loops $\vec{n}$, here choosen
to be a random integer vector \cite{Deleo}. $\hat{K}$ and $\hat{L}^{-1}$ are the matrixes carrying the mutual inductance coupling. The vector $\vec{\psi}$ has values of $0$ or $\pi$, providing a $\pi$-shift to a subset of the junctions in the array. The concentration $c$ of 
$\pi$-junction \cite{Dominguez} is the ratio between the number of $\pi$-junctions and the 
total number of junctions. The vector $\vec{\delta}$ is a Gaussian variable with zero mean 
which we use to introduce disorder in the distribution of critical currents. 

Here we suppose that in the array the contribution of the screening currents is not negligible, i.e., $\beta_L>1$ (cf. \cite{AltriHTC}). In particular we choose $\beta_L=30$ and $\beta_C=63$ (cf. refs. \cite{Nielsen,Deleo}).

In both experiments \cite{Nielsen} and previous simulations \cite{Deleo} it was shown that multiple-connectiveness effects are dominant for large field, i.e., for large values of frustation $f$. To understand the effect of the $\pi$-shift we studied smaller values of 
frustation $0<f<1$. In this region the behavior of an array without disorder and with all normal or all $\pi$-junctions roughly follows the single loop behavior \cite{Nielsen,Deleo,Moreira}which is described by
\begin{equation}
\frac{\Phi}{\Phi_0}-f
=\frac{\beta _{L}}{2\pi }\;
\sin (\frac{\pi}{2}n-\frac{\psi}{4}-\frac{\Phi}{\Phi_0})
\label{1loop}
\end{equation}
where $\Phi$ is the total flux throught the loop. Solution of Eq. (2) gives the possible
states for the total flux $\Phi$ (or the current) in the single loop.
For $\psi=0$ Eq. (2) describe a 
single loop made by four conventional Josephson junction. There are four independent solutions which are obtained varying the integer $n$ from zero to three. The (stable) lowest Gibbs energy state is diamagnetic for $0<f<1/2$, i.e., the loop current is negative, generating a magnetic moment opposite to the external field, and paramagnetic for $1/2<f<1$, i.e., the loop current is positive \cite{Nielsen,Deleo}. On the other hand if $\psi=\pi$ the $\pi$-shift will reverse the behavior: for the $\pi$-loop (a loop with
an {\it odd} number of $\pi$-junctions) the lowest Gibbs energy state is paramagnetic for $0<f<1/2$ and diamagnetic for $1/2<f<1$.

In Ref. \cite{Deleo} we showed that conventional-array behavior can be qualitatively understood
in terms of a model based on the single-loop behavior. We found that each loop in the array had a magnetization that was close to one of the lowest energy solutions of Eq. (2) for the
isolated loop, with the number of loops of each type determining the overall magnetization.
This simple picture also works for arrays with both conventional and $\pi$-junctions,
with the number of possible solutions being increased because that Eq. (2) has
additional solutions for the $\pi$-loops.

In order to compare conventional arrays with mixed conventional/$\pi$ arrays, we first
simulate the conventional arrays using the method of Ref. \cite{Deleo}. In Fig. 2a the magnetization of a $20\times 20$ conventional array 
with $f=0.2$ is shown. There are two single-loop magnetization states with a predominancy
of diamagnetic states (light gray) over which a set of paramagnetic loops (dark gray) are randomly distributed. The histogram of loop magnetizations is shown in
Fig. 3a and shows clearly only two peaks corresponding to two loop
magnetizations. The mean magnetization is negative and small as 
expected for such small values of frustation accordingly to the Eq. (2) for the single loop; 
the exact value is $m=-0.0027$. As said above the values of the 
single peak averages can be obtained from Eq. (2); the
small spread of the peaks is due to mutual inductance coupling.

In Fig. 2b a similar $20\times 20$ array with $f=0.2$ is shown in which $380$ $\pi$-junctions have been introduced at random. This is equivalent to 
a concentration of $\pi$-junctions of $c=0.45$. The main 
difference with the case of coventional array is the presence of four 
magnetization states, two for each type of loop. 
We note that the new $\pi$-loop states are predominantly paramagnetic according to the
symmetry shift discussed above. This, together with the lower number of
conventional loops, is causing the reversal of the mean magnetization (here being $m=+0.0018$). This is the way in which a mixed array becomes paramagnetic for low values of frustation. 
Lower concentrations will make the array diamagnetic (i.e. the dependence on concentration is similar to that reported in Fig. 3b of Ref. \cite{Dominguez} for a 3D system). For other values of $f$ we note that the single loop symmetry
is broken, i.e., for $f<1/2$ the array can be diamagnetic or paramagnetic depending
on the distribution and number of $\pi$-loops. 

From the preceding discussion, it is clear that there are at best only subtle
differences between a conventional array and an array containing a mixture of
conventional and $\pi$-junctions. In the remainder of the paper we will show that
when $f=1/2$ these differences are the most pronounced, and therefore $f=1/2$ is
probably the best place to look experimentally in searching for evidence of
$\pi$-junctions in non zero magnetic field. 
When $f=1/2$ the problem is simplified by the fact that the four lowest states shrink to
three because the $\pi$-loop magnetization becomes zero.
So one is confronted with only three peaks which are symmetrically
distributed around the zero magnetization. In Fig. 4a this case is
shown in a magnetization histogram. The central peak is the zero magnetization 
of $\pi$-loops and two symmetric lateral peaks are due to coventional loops. 
Symmetrical histogram structure is an optimal "signature", it is easily recognized with 
SSM and statistical filtering or image filtering techniques can be applied knowing the true 
distribution \cite{Wikswo}.
 
The central peak intensity is related to presence of $\pi$-loops.
At $c=0.3$ ($250$ $\pi$-junctions) it becomes thinner preserving the same
height. Then it decreases by about 10\% at $c=0.15$ ($126$ $\pi$-junctions).
Finally at $c=0.05$ ($42$ $\pi$-junctions) it decreases of about 60\%.   

In Fig. 4a a weak Gaussian disorder with a standard deviation of 
$\sigma=20\%$ was introduced 
in the critical currents of the arrays similarly to Ref.\cite{Disorder}.  
This was added to shown that weak disorder does not change the three-peak
structure of the magnetization histogram. For high values of the parameter 
$\beta_L$ the nature of the magnetization states is discrete and this is 
preserved for weak disorder in the currents. Peak structure exists also with 
stronger disorder in the critical currents. In Fig. 4b we set $\sigma=80\%$\cite{nota}
we see peaks enlarging until the tails of the single three distributions 
overlaps. It is observed that the remaining discrete structure is due again to the 
large $\beta_L$ loops. Also we have simulated the array 
with the same parameters of Fig. 4b using a mean value of $1/3$ for the $\pi$-junction 
critical current. The result is again similar to Fig. 4b but that central peak is 
lowered by roughly 13\% and the histogram appears more enlarged.

In conclusion we would stress again how both conventional and mixed conventional/$\pi$
arrays show a similar response when field-cooled. This response
is related to the single loop solutions of Eq. (2). In low field $f<0.5$ paramagnetism in
mixed conventional/$\pi$ arrays is the result of the paramagnetic behavior
of $\pi$-loops. To quantify this effect for a disorderd system is difficult
because it depends on distribution and number of $\pi$-loops.
On the other hand a noteworthy point is $f=1/2$. Here the $\pi$-loops sets to 
zero current so it is possible from a measurement of the magnetization histogram to
trace back the presence of $\pi$-loops in the sample. For High-T$_c$ materials this 
experiment can be an important complement of the search for spontaneous currents in zero field \cite{Tafuri}. Correlation between $f=0$ and $f=1/2$ would be of great interest permitting 
to estimate the $\pi$-loops content of the sample.  
  
We warmly thank P. Barbara and C. J. Lobb for useful discussions and suggestion. 
Our thanks would be given also to A. P. Nielsen, F. Tafuri, G. Filatrella, S. Pagano and
G. Costabile for their useful comments. We 
acknowledge financial support from MIUR\ COFIN2000 project 
''Dynamics and Thermodynamics of vortex structures in superconductive tunneling''.

\newpage

\section*{Figure Captions}
Fig. 1 Two-dimensional Josephson junction array with a random distribution of $\pi$-junctions. The gray loops represent the $\pi$-loops, i.e., loops with an odd number of $\pi$-junctions.
The figure is just the top-left corner of the mixed array in Fig. 2b.

Fig. 2 Simulated magnetization of a $20\times 20$ array at $f=0.2$ with $\beta_L=30$ and $\beta_C=63$: a) without $\pi$-junctions, light gray represents diamagnetic loops, gray paramagnetic ones; b) with $\pi$-junctions at a concentration of $c=0.45$ white and light 
gray represents diamagnetic loops, gray and dark gray paramagnetic ones. 

Fig. 3 Magnetization histogram of a $20\times 20$ plane array at $f=0.2$ with $\beta_L=30$ 
and $\beta_C=63$: a) distribution without $\pi$-junctions; b) distribution with 
$\pi$-junctions at a concentration of $c=0.45$. 

Fig. 4 Magnetization histogram of a $20\times 20$ plane array at $f=0.5$ with $\beta_L=30$, $\beta_C=63$ and $\pi$-junctions distribution at $c=0.45$: a) effect of a weak disorder in the current distribution ($\sigma=20\%$); 
b) the same for a stronger disorder($\sigma=80\%$).


\begin{thebibliography}{99}
\bibitem{nota} Note that points in the tails of Gaussian distribution of the currents can pick out negative values, so reverting some $\pi$-junction in normal junction and viceversa.
Statistical tests shown that the number is however again balanced around $c=0.45$.
\end{thebibliography}
\end{document}